\documentclass[aip, reprint]{revtex4-1}
\usepackage{hyperref}
\usepackage{graphicx}
\usepackage{upgreek}

\draft 

\begin{document}


\title{Superconducting properties of Ba(Fe$_{1-x}$Ni$_x$)$_2$As$_2$ thin films in high magnetic fields} 

\author{Stefan Richter}
 \email{s.richter@ifw-dresden.de}
 \affiliation{ 
 Institute for Metallic Materials, IFW Dresden, 01171 Dresden, Germany
 }%
 \affiliation{ 
  TU Dresden, 01062 Dresden, Germany
  }%
\author{Fritz Kurth}%
\affiliation{ 
Institute for Metallic Materials, IFW Dresden, 01171 Dresden, Germany
}%

\author{Kazumasa Iida}
\affiliation{%
Department of Crystalline Materials Science, Nagoya University, Chikusa, Nagoya 464-8603, Japan
}%

\author{Kirill Pervakov}
\affiliation{%
Department of High Temperature Superconductivity and Superconductor Nanostructures, Russian Academy of Sciences, 119991 Moscow, Russia
}%

\author{Aurimas Pukenas}%
\affiliation{ 
TU Dresden, 01062 Dresden, Germany
}%

\author{Chiara Tarantini}
\affiliation{%
Applied Superconductivity Center, National High Magnetic Field Laboratory, Florida State University, 2031 East Paul Dirac Drive, Tallahassee, Florida 32310, USA
}%

\author{Jan Jaroszynski}
\affiliation{%
Applied Superconductivity Center, National High Magnetic Field Laboratory, Florida State University, 2031 East Paul Dirac Drive, Tallahassee, Florida 32310, USA
}%

\author{Jens H\"{a}nisch}%
\affiliation{ 
Institute for Technical Physics, Karlsruhe Institute of Technology, Hermann-von-Helmholtz-Platz 1, 76344 Eggenstein-Leopoldshafen, Germany 
}%

\author{Vadim Grinenko}
\affiliation{ 
Institute for Metallic Materials, IFW Dresden, 01171 Dresden, Germany
}%
\affiliation{ 
TU Dresden, 01062 Dresden, Germany
  }%
  
\author{Werner Skrotzki}%
\affiliation{ 
TU Dresden, 01062 Dresden, Germany
}%

\author{Kornelius Nielsch}%
\affiliation{ 
Institute for Metallic Materials, IFW Dresden, 01171 Dresden, Germany
}%
 \affiliation{ 
  TU Dresden, 01062 Dresden, Germany
  }%
  
\author{Ruben H\"{u}hne}%
\affiliation{ 
Institute for Metallic Materials, IFW Dresden, 01171 Dresden, Germany
}%


\date{\today}

\begin{abstract}
We report on electrical transport properties of epitaxial Ba(Fe$_{1-x}$Ni$_x$)$_2$As$_2$ thin films grown by pulsed laser deposition in static magnetic fields up to 35\,T. The thin film shows a critical temperature of 17.2\,K and a critical current density of 5.7\,$\times$\,10$^5$\,A/cm$^2$ in self field at 4.2\,K while the pinning is dominated by elastic pinning at two-dimensional nonmagnetic defects. 
Compared to single-crystal data, we find a higher slope of the upper critical field for the thin film at a similar doping level and a small anisotropy. 
Also an unusual small vortex liquid phase was observed at low temperatures, which is a striking difference to Co-doped BaFe$_2$As$_2$ thin films.   
\end{abstract}


\maketitle
Fe-based superconductors are intriguing materials for both basic research and potential applications. In particular, the electronic phase diagrams show competing ground states, which lead to unique normal state properties and unconventional superconductivity \citep{Hirsch}. 
These materials also have the potential to supersede low-temperature superconductors in high-field applications since they combine the advantage of high upper critical fields with a small anisotropy at low temperatures \cite{potential}. Also a higher tolerance to low angle grain boundaries compared to cuprates has been found for Co-doped BaFe$_2$As$_2$ (Ba122) \citep{grainboundary}. 
Ba122 is among the most studied systems of the Fe-based superconductors since single crystals and thin films are relatively easy to grow and several kinds of doping (hole, electron, isovalent, direct/indirect) lead to superconductivity \citep{doping_growth}. We used Ni as dopant \citep{first_Nickel}, which has two electrons more in the $d$ shell than Fe and will therefore dope electrons in the system similar to the extensively studied Co substitution. 
Ni-doped Ba122 was found to be an interesting candidate for applications due to inherently strong pinning effects \citep{Kirill} leading to a high critical current density $J_\mathrm{c}$. 

Here we report on thin films of Ba(Fe$_{1-x}$Ni$_x$)$_2$As$_2$ with a thickness of 100\,nm which have been prepared by pulsed laser deposition (PLD) using a KrF excimer laser ($\lambda$\,=\,248\,nm, pulse duration 25\,ns, energy density 3\,J/cm$^2$, repetition rate 7\,Hz). The deposition was performed under ultra high vacuum conditions (base pressure 10$^{-9}$\,mbar) at a deposition temperature of 750\,$^\circ$C. A polycrystalline pellet was prepared from a nominal Ba(Fe$_{0.95}$Ni$_{0.05}$)$_2$As$_2$ powder and served as PLD target. The Ni-content of the final target was determined to be in the overdoped regime ($x$\,=\,0.065\,$\pm$\,0.015) using energy dispersive X-ray spectroscopy (EDX) analysis in a JEOL scanning electron microscope.
CaF$_2$ (001) was used as substrate, which has been proven as a suitable template for the growth of Ba122 thin films \cite{flouride}.

\begin{figure}[t]
\centering
\includegraphics[width=\columnwidth]{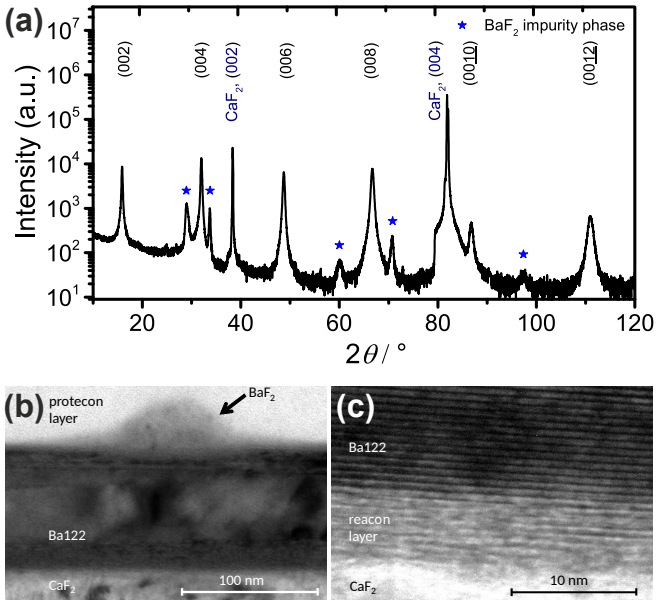}
\caption{Microstructural properties of the investigated Ni-doped Ba122 film: a) XRD 2$\theta$ scan showing the $c$-axis oriented growth of the Ba122. A detected BaF$_2$ impurity phase is marked with blue stars. 
b) In TEM, BaF$_2$ impurities are found as precipitates on the film surface (black arrow). 
c) A reaction layer of 5-10\,nm thickness at the film-substrate interface was observed. The Ba122 phase above this reaction layer shows a good crystalline quality without any observable defects.}
\label{xrd2}
\end{figure}

X-ray diffraction (XRD) using a Bruker D8 Advance diffractometer with Co-K$_\alpha$ radiation shows only (00$l$) reflections of the Ba122 and the substrate (Figure \ref{xrd2}a) indicating a $c$-axis oriented film, together with peaks corresponding to a BaF$_2$ impurity  phase.

The distribution of this BaF$_2$ phase was investigated by transmission electron microscopy (TEM) on a cross sectional lamella along the [100] direction (prepared with a focused ion beam technique) using an FEI Tecnai T20  microscope (accelerator voltage\,=\,200\,kV). Precipitates with a lateral size of about 150\,nm were found primarily on the film surface (Figure \ref{xrd2}b), i.e. not incorporated in the superconducting layer. Selected area diffraction confirmed them as BaF$_2$. The BaF$_2$ is formed presumably due to a reaction of excess Ba with fluorine diffusing out of the substrate. 
Besides these precipitates, the film shows a clean microstructure. Planar defects, which were reported occasionally for Co-doped Ba122 PLD films \citep{jens,haindl,Trommler} are almost absent in our film. A partly amorphous reaction layer at the film-substrate interface of about 5-10\,nm thickness was observed (Figure \ref{xrd2}c) similar to Co-doped Ba122 films on CaF$_2$ \citep{flouride}. It is likely caused by beam damage during the lamella preparation. 
XRD texture measurements (data not shown) reveal the epitaxial relation (001)[110]Ba122$\parallel$(001)[100]CaF$_2$. The full width at half maximum (FWHM) of the (103) $\phi$ scan ($\Delta\phi$\,=\,0.93$^\circ$) and of the (004) rocking curve ($\Delta\omega$\,=\,0.71$^\circ$) (both values not corrected for instrument broadening and geometrical effects) indicate a good crystalline quality of the superconducting layer and are comparable to the results for Co-doped Ba122 thin films on CaF$_2$ \citep{flouride}. 
For the determination of the $c$-axis length, the peak positions of the 2$\theta$ scan and the Nelson Riley extrapolation \citep{nelsonriley} were used, while the $a$-axis was extracted from  reciprocal space maps measured in a high resolution X-ray diffractometer. With lattice parameters $c$\,=\,13.031$\,\pm$\,0.002\,\AA\, and $a$\,=\,3.949$\,\pm$\,0.001\,\AA, the Ba122 unit cell is elongated along the $c$-direction in the thin film in comparison to single crystals with similar doping ($c$\,=\,12.986\,\AA\, $a$\,=\,3.963\,\AA\, for $x$\,=\,0.065) \citep{scgrowth, doping_lattice_parameters}. This originates from compressive strain in the film due to the smaller $a$-axis parameter and larger thermal expansion coefficient of the CaF$_2$ substrate\citep{Lei_strain, kazu_strain}.

\begin{figure}[t]
\centering
\includegraphics[width=\columnwidth]{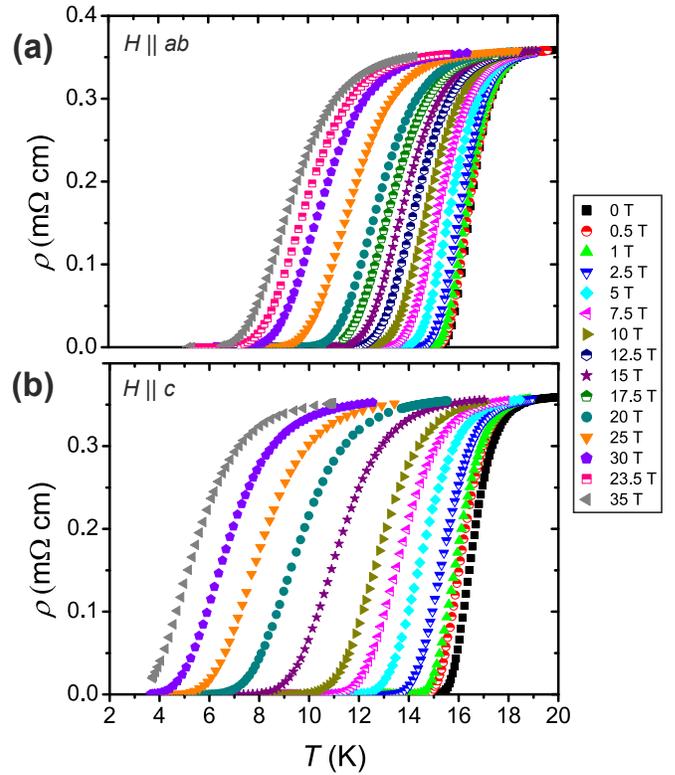}
\caption{Superconducting transition in applied magnetic fields for: a) $H\parallel ab$ and b) $H\parallel c$.}
\label{R(T)_2}
\end{figure}

Transport properties have been measured using a four-point technique on a bridge of 2\,mm length and 90\,$\upmu$m width prepared by laser cutting. Measurements were performed in static magnetic fields up to 35\,T in maximum Lorentz force configuration at the National High Magnetic Field Laboratory. The critical temperature $T_\mathrm{c}$, evaluated at 90\,\% of the normal resistance is 17.3\,K (Fig.\,\ref{R(T)_2}). The superconducting transition is sharp with $\Delta T_\mathrm{c}$\,=\,$T_\mathrm{c90}$\,-\,$T_\mathrm{c10}$\,\textless\,1\,K.The broadening of the transition with increasing external field is relatively small, which is commonly observed in other Ba122 systems due to a small Ginzburg number. 
The observed $T_\mathrm{c}$ value is lower than results on optimally doped single crystals \citep{wang} but fits well to data for slightly overdoped material in agreement with the measured Ni-content of the target. However, the direct comparison of $T_\mathrm{c}$ values between single crystals and thin films might be problematic due to complex effects of strain on the phase diagram. It has been shown for Co-doped Ba122, that compressive strain can even lead to higher transition temperatures in slightly overdoped thin films compared to optimally doped single crystals \citep{kazu_strain}.
 
\begin{figure}[t]
\centering
\includegraphics[width=\columnwidth]{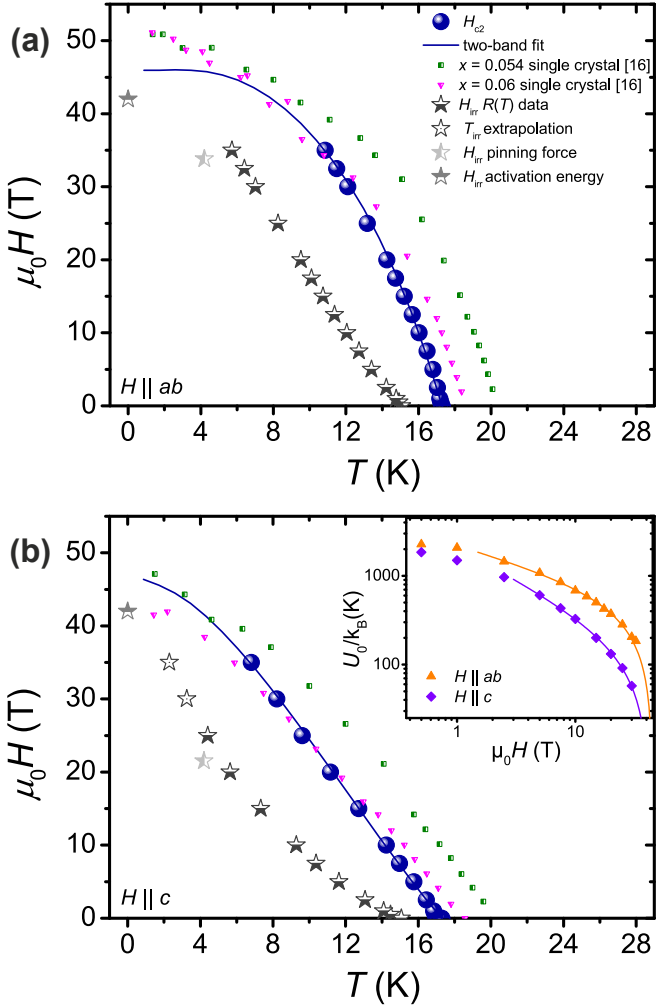}
\caption
{Upper critical field and irreversibility field data for a) $H$$\parallel$$ab$ and b) $H$$\parallel$$c$. Blue dots show $H_\mathrm{c2}$ using a 90\% criterion and the blue lines represent two-band fits of these data. For comparison, results from Wang et al. \citep{wang} for single crystals are shown for $x$\,=\,0.054 (green squares) and $x$\,=\,0.06 (magenta triangels). The stars show $H_\mathrm{irr}$ determined from the resistivity data (dark grey), the fit of the pinning force density (light grey) and the fit of the activation energy (grey). Open symbols are extrapolated from Arrhenius plots. 
The inset shows the activation energy for both field orientations. The solid lines are fits using eq. (\ref{U_0}) \citep{activation}.}
\label{critical_fields}
\end{figure}

A 90\,\% criterion at 17.3\,K was used for the evaluation of the upper critical field $H_\mathrm{c_2}$ and the irreversibility field $H_\mathrm{irr}$ was evaluated at $\rho$\,=\,10$^{-4}$\,m$\Omega$cm (which is roughly 0.5\% of the normal state resistivity at 30\,K). The data shown in Figure \ref{critical_fields} are in good agreement with recent pulsed field measurements for overdoped single crystals \citep{wang}.
The $H_\mathrm{c_2}$ data were fitted using a two-band model for a clean superconductor including paramagnetic effects \citep{Two_band_model_Hc2}. 
The fit suggests an upper critical field of 45\,T and a smaller anisotropy than in overdoped single crystals with a similar critical temperature \citep{wang}. The slope of the upper critical field \mbox{-d$H_\mathrm{c2}$/d$T$\,=\,8.5\,T/K} ($H$$\parallel$$ab$) near $T_\mathrm{c}$ is higher for the film than for single crystals with similar doping level and more comparable to optimally doped crystals \citep{wang}. 
This might be explained with the presence of epitaxial strain in the film
leading to a change in the band structure, which results in smaller Fermi velocities \citep{kazu_strain, Chiara}. 
Interestingly, at low temperatures the irreversibility field almost seems to reach the $H_\mathrm{c2}$ line as its natural limitation for both field directions. This would mean that Ba(Fe$_{1-x}$Ni$_x$)$_2$As$_2$ thin films show a small vortex liquid phase at low temperatures in contrast to results for Co-doped films \citep{jens}.
However, one should take into account that $H_\mathrm{c2}$ values at low temperatures could be additionally enhanced due to the formation of the Fulde-Ferrel-Larkin-Ovchinnikov (FFLO) state \citep{Two_band_model_Hc2}. This possibility was not taken into account in the $H_\mathrm{c2}$ analysis in the present work due to a lack of experimental data at low temperatures. Future investigations in high fields will clarify this question. 

\begin{figure}[t]
\centering
\includegraphics[width=\columnwidth]{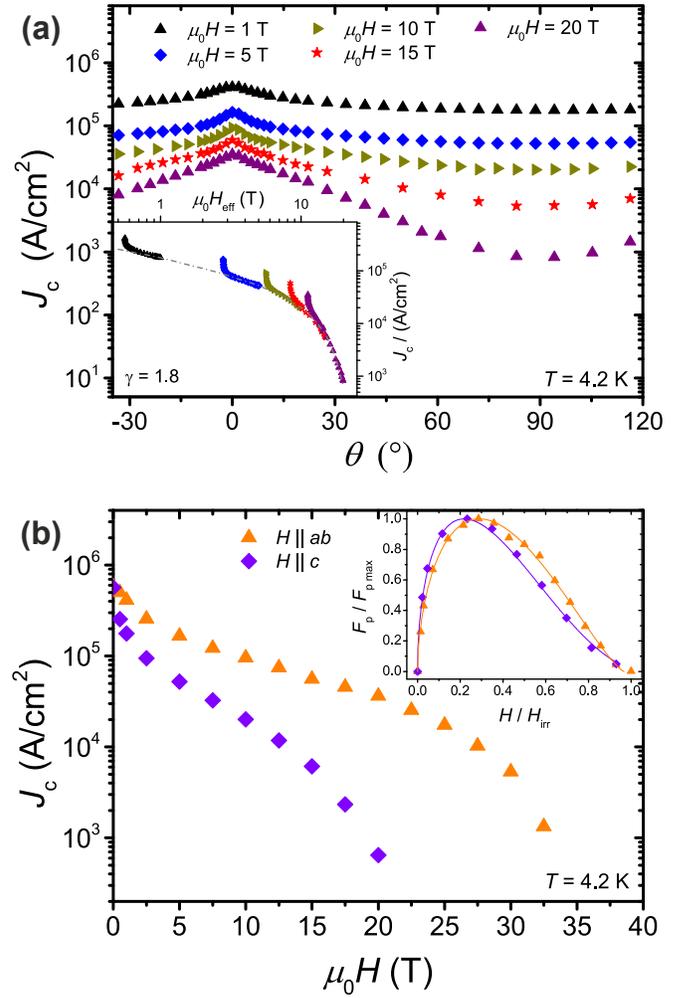}
\caption
{
 Critical current density for 4.2\,K. a) The angular dependence of $J_\mathrm{c}$ shows a maximum for $H$$\parallel$$ab$. The inset shows the scaling over an effective field (\ref{Heff}) \citep{Blatter}. b) Magnetic field dependance for $H$$\parallel$$ab$ (orange triangels) and $H$$\parallel$$c$ (purple diamonds). The inset shows the normalized pinning force density and fits using equation (\ref{Fp}).
 }
\label{j_c}
\end{figure}

The linear dependence of Arrhenius plots of the resistivity (not shown here) indicates thermally activated motion of the flux lines above $H_\mathrm{irr}$. The pinning potential $U_0$ extracted from the Arrhenius slopes is shown in the inset of figure \ref{critical_fields}b.  The solid lines represent fits with the phenomenological function \citep{activation, jens}:

\begin{equation}
U_0 \propto H^{-a}\left( 1- \frac{H}{H_\mathrm{irr}} \right)^b  
\label{U_0}
\end{equation}

We achieved the best fits with parameters $a$\,=\,0.4, $b$\,=\,1.0, $\mu_\mathrm{0}$$H_\mathrm{irr}$\,=\,45\,T for $H$$\parallel$$ab$ and 
$a$\,=\,0.7, $b$\,=\,1.0, $\mu_\mathrm{0}$$H_\mathrm{irr}$\,=\,42\,T for $H$$\parallel$$c$.

The critical current density $J_\mathrm{c}$ is shown in figure \ref{j_c}. The $J_\mathrm{c}$ at 4.2\,K in self field exceeds 5.7\,$\times$\,10$^5$\,A/cm$^2$. This is lower than $J_\mathrm{c}$ values for differently doped Ba122 films \citep{highjc, P_doped}, presumably due to the lower $T_\mathrm{c}$. 
Also magnetisation measurements on optimally doped single crystals \citep{Kirill} showed a higher $J_\mathrm{c}$ while it was highest for $H$$\parallel$$c$, i.e. an inversion of the $J_\mathrm{c}$-anisotropy occured. 
This inversion was not observed for our thin films. 

Figure \ref{j_c}a shows the angular dependance of $J_\mathrm{c}$ at $T$\,=\,4.2\,K. Only one maximum for $H$$\parallel$$ab$ is observed. The data can be scaled over an effective field \citep{Blatter}: 

\begin{equation}
H_\mathrm{eff} = H\cdot \sqrt{\sin^2(\theta)+\frac{\cos^2(\theta)}{\gamma^2}}
\label{Heff}
\end{equation}

An anisotropy value of $\gamma$\,=\,1.8 was found for the critical current density at 4.2\,K.

The field dependence of $J_\mathrm{c}$ is shown in figure \ref{j_c} b). The inset shows the normalized pinning force density $F_\mathrm{p}$, which was fitted with the Dew Hughes model \citep{DewHughes}: 
\begin{equation}
F_\mathrm{p} \propto \left( \frac{H}{H_\mathrm{irr}} \right)^p \left( 1- \frac{H}{H_\mathrm{irr}}\right)^q
\label{Fp}
\end{equation}

\begin{figure}[t]
\centering
\includegraphics[width=\columnwidth]{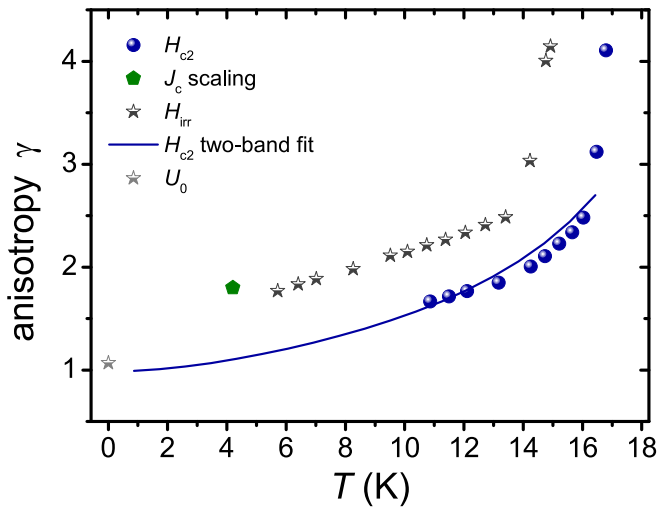}
\caption
{The upper critical field and the activation energy show a small anisotropy close to 1 at low temperatures. In comparison, the anisotropy of the critical current density and the irreversibility field is noticeably increased.}
\label{gamma}
\end{figure}

The respective fit parameters for $H$$\parallel$$ab$ and $H$$\parallel$$c$ are $p$\,=\,0.56, $q$\,=\,1.32 and $p$\,=\,0.48, $q$\,=\,1.67. The corresponding $H_\mathrm{irr}$ values are shown in figure \ref{critical_fields}. 
The $p$-values of the pinning force fits for both field directions suggest, that the pinning is influenced by elastic effects of the vortex lattice and vortex interactions. This contribution to the pinning is significant, if the size of the pinning centers $r$ is smaller than or comparable to the coherence length $\xi$\, \citep{pinning_diameter}. For $H$$\parallel$$c$, the fit is close to ($p$\,=\,0.5, $q$\,=\,2), which can be attributed to pinning at two-dimensional nonmagnetic defects \citep{DewHughes}. In our film these defects could be growth domain boundaries or small-angle grain boundaries, although they have not been directly confirmed in TEM investigations. 
For $H$$\parallel$$ab$, the exponent $q$ is smaller, which has been observed before in Co-doped \citep{jens} and P-doped \citep{P_doped} Ba122 films at low temperatures. This might be related to the small film thickness, which restricts the bending of vortices. For $H$$\parallel$$ab$, the smaller $q$ value is in qualitative agreement with the results of the activation energy, since the exponents of the activation energy (\ref{U_0}) and the pinning force density (\ref{Fp}) should fulfill the relations $a\approx 1-p$, $b\approx q$ \citep{activation}. 
For $H$$\parallel$$c$ these equations are not fulfilled, which has also been observed for Co-doped films \citep{jens}.

The anisotropy $\gamma$ of the critical fields and critical current density is shown in figure \ref{gamma}. For $H_\mathrm{c2}$ and $H_\mathrm{irr}$ it is calculated by $\gamma\,=\,\frac{H^{ab}}{H^c}$ where $H^{ab}$ and $H^c$ represent the critical field for the respective field orientation. 
The data points and the two-band fit for $H_\mathrm{c2}$ suggest a small anisotropy close to 1 at low temperatures. This agrees well with the anisotropy extracted from the activation energy fits. Such a small anisotropy despite of the highly anisotropic crystal structure is a key feature of iron-based superconductors \citep{Zhang_gamma}.
The $J_\mathrm{c}$ anisotropy follows the $H_\mathrm{irr}$ anisotropy and both are noticeably increased compared to the $H_\mathrm{c2}$ anisotropy. This indicates that the incorporated defects in the film affect the $H_\mathrm{irr}$ anisotropy, differentiating it from the intrinsic $H_\mathrm{c2}$ anisotropy. 

The enhancement of the $H_\mathrm{irr}$ anisotropy is typically observed in Fe-based superconductors by introduction of pinning centers \citep{jens,feifei}, which is in contrast to cuprate high-$T_\mathrm{c}$ superconductors, where the addition of nanoparticles often leeds to a reduction in the $H_\mathrm{irr}$ and $J_\mathrm{c}$-scaling anisotropy \citep{YBCO_gamma}.

In summary, we reported the epitaxial growth of Ba(Fe$_{1-x}$Ni$_x$)$_2$As$_2$ thin films, investigated the structural and superconducting properties and evaluated their performance in high magnetic fields. The films show a $T_\mathrm{c}$ value of about 17\,K and an enhanced slope of the upper critical field at $T_\mathrm{c}$ in comparison to single crystals with similar doping level. The enhanced slope \mbox{-d$H_\mathrm{c2}$/d$T$} might be explained by the presence of compressive strain in the film. The critical current density reaches 5.7\,$\times$\,10$^5$\,A/cm$^2$ at 4.2\,K in self field which is smaller than the best results achieved for Co-doped Ba122 thin films \citep{highjc}.
The pinning is dominated by elastic pinning at grain boundaries or growth domain boundaries and the anisotropy at low temperatures is small similar to results for other Fe based superconductors.
Furthermore, we found the vortex liquid region of the vortex-matter phase diagram for Ni-doped films to be rather small which is surprisingly different to Co-doped Ba122 films.

This work was supported by the German research foundation (DFG) within the research training group GRK 1621.
A portion of this work was performed at the National High Magnetic Field Laboratory, which is supported by National Science Foundation Cooperative Agreement No. DMR-1157490 and the State of Florida. This work was also partially supported by JSPS Grant-in-Aid for Scientific Research (B) grant number 16H04646 and the DFG grant number GR4667/1-1.

\end{document}